\documentclass[letterpaper]{article} 
\usepackage{aaai2026}
\usepackage{times}  
\usepackage{helvet}  
\usepackage{courier}  
\usepackage[hyphens]{url}  
\usepackage{graphicx} 
\usepackage{natbib}  
\usepackage{caption} 
\usepackage{algorithm}
\usepackage{algorithmic}

\usepackage{newfloat}
\usepackage{listings}
\DeclareCaptionStyle{ruled}{labelfont=normalfont,labelsep=colon,strut=off} 
\floatstyle{ruled}
\newfloat{listing}{tb}{lst}{}
\floatname{listing}{Listing}
\usepackage{booktabs}
\usepackage{graphicx}
\usepackage{amsmath,amssymb}
\usepackage{multirow}

\usepackage{xcolor}
\newcommand{\answerYes}[1]{\textcolor{blue}{#1}} 
 
\newcommand{\answerNA}[1]{\textcolor{gray}{#1}} 
 
\usepackage{booktabs}

\usepackage[caption=false]{subfig}

\title{The Gray Area: Characterizing Moderator Disagreement on Reddit}

\author{
    Shayan Alipour\textsuperscript{\rm 1,4},
    Shruti Phadke\textsuperscript{\rm 2},
    Seyed Shahabeddin Mousavi\textsuperscript{\rm 3},
    Amirhossein Afsharrad\textsuperscript{\rm 3},
    Morteza Zihayat\textsuperscript{\rm 4},
    Mattia Samory\textsuperscript{\rm 1}
} 

\affiliations{
    \textsuperscript{\rm 1} Sapienza University of Rome\\
    \textsuperscript{\rm 2} Drexel University\\
    \textsuperscript{\rm 3} Stanford University\\
    \textsuperscript{\rm 4} Toronto Metropolitan University\\
    shayan.alipour@uniroma1.it, shruti.phadke@drexel.edu,
    ssmousav@cs.stanford.edu,
    afsharrad@stanford.edu,
    mzihayat@torontomu.ca,
    mattia.samory@uniroma1.it
}

\begin{document}

\maketitle

\begin{abstract}
Volunteer moderators play a crucial role in sustaining online dialogue, but they often disagree about what should or should not be allowed. In this paper, we study the complexity of content moderation with a focus on disagreements between moderators, which we term the ``gray area'' of moderation. Leveraging 5 years and 4.3 million moderation log entries from 24 subreddits of different topics and sizes, we characterize how gray area, or disputed cases, differ from undisputed cases. 
We show that one-in-seven moderation cases are disputed among moderators, often addressing transgressions where users' intent is not directly legible, such as in trolling and brigading, as well as tensions around community governance. This is concerning, as almost half of all gray area cases involved automated moderation decisions. Through information-theoretic evaluations, we demonstrate that gray area cases are inherently harder to adjudicate than undisputed cases and show that state-of-the-art language models struggle to adjudicate them. We highlight the key role of expert human moderators in overseeing the moderation process and provide insights about the challenges of current moderation processes and tools.
\end{abstract}

\begin{figure*}[htbp]
    \centering
    \includegraphics[width=0.92\linewidth]{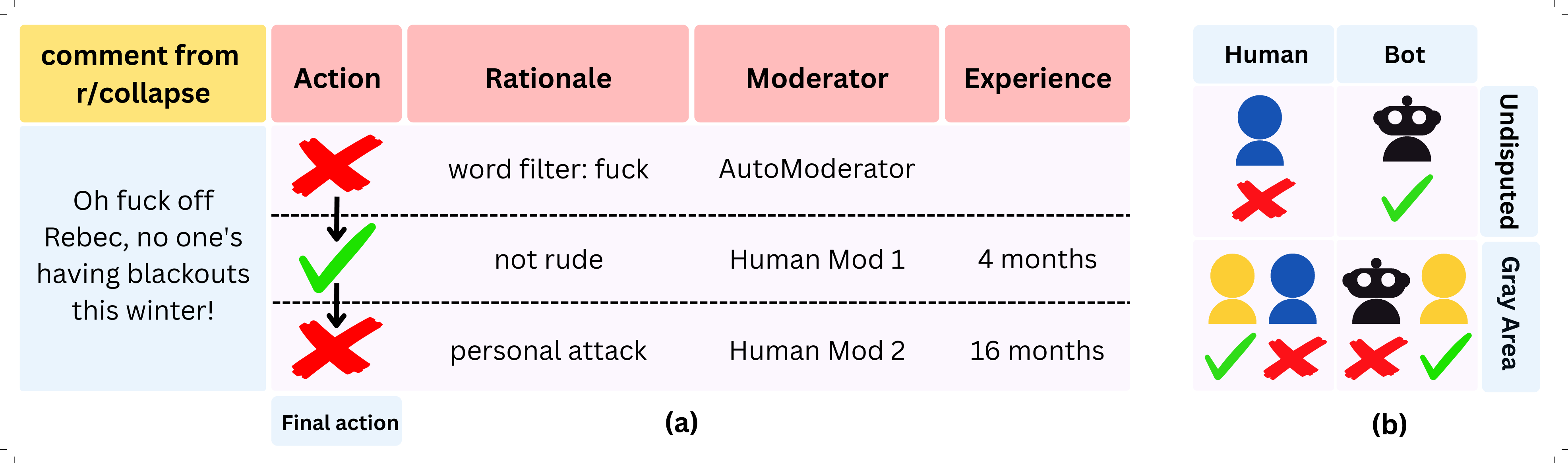}
    \caption{(a) An example showing multiple moderation actions on a single comment. (b) Cases are partitioned into four mutually exclusive strata by (i) whether there is within-case disagreement defined as more than one unique moderator and more than one unique action and (ii) whether any moderator is a bot. The case (a) is categorized into ``gray bot''.}
    \label{fig:worked_example}
\end{figure*}

\section{Introduction}\label{sec:intro}

Online communities require moderation to function, with content decisions determining which voices participate in public discourse. 
Volunteer moderators handle the vast majority of content decisions on Reddit, providing labor that rivals professional staff in scale~\cite{li2022all}. 
Their work involves evaluating millions of posts and comments against community norms, often under time pressure and with limited resources. Regardless of their decisions, the moderators' role exposes them to user criticism, creating burnout within volunteer moderation teams \cite{schpke-gonzalez2022why-2e8}. 

Gray area cases, which require disputation between moderators before adjudication, embody a key 
tension between strained moderator resources and the deliberative needs of community governance.
Disagreement indicates that moderators engage with boundary-setting, which fosters community growth by establishing precedents and reshaping norms. 
Studies of moderation practices acknowledge the necessity of effective disagreement management to improve governance health~\cite{jhaver2019did-151,vaccaro2021contestability-93d}. Research demonstrates that inherent ambiguity in content interpretation, rule application, and contextual and subjective factors may create legitimate disagreement among moderators~\cite{doi:10.1177/14614448231190901}. Design approaches to participatory moderation have, therefore, emerged to accommodate multiple viewpoints rather than forcing artificial consensus, valuing disagreement~\cite{falk2024moderation-ee8,lam2022end-user-6b2,gordon2022jury-3b9}. 


Despite the importance of disputed cases, we lack an empirical understanding of the extent and nature of the gray area within existing moderation systems. The opacity of moderation processes makes it difficult to systematically evaluate the factors that drive disagreement and assess how communities adjudicate them. 
This has implications for the fairness of the moderation process, since the apparent discretion of moderator decisions, especially when subject to dispute, may disproportionately affect vulnerable populations and erode support of the process \cite{haimson2021disproportionate-26f,juneja2020through}.

This work addresses these gaps through analysis of a longitudinal and complete dataset of moderation decisions in 24 subreddits participating in open moderation logging, collected over 5 years. 
We characterize the gray area by comparing disputed cases against undisputed moderation decisions, disaggregated by human versus automated moderator involvement. 
First, through statistical analyses, manual coding, and topic modeling, we find that the gray area is common (about one in seven cases), that many disputes arise around ambiguous or sensitive content (e.g., trolling, brigading, hate/harassment), and that automated removals are frequently reversed by humans, especially as moderator experience increases. 
Next, we assess the viability of LLM-based moderation for gray area cases through in-context learning and information-theoretic analyses, finding that gray cases are intrinsically harder and that current models underperform on disputed content relative to undisputed cases. 

Finally, we show that LLM--moderator alignment on gray area cases is driven both by local community context and the inherent complexity of the text, while being significantly more consistent for removal decisions than for approvals.

Highlighting a key component of moderation processes, this work adds to a growing body of research documenting its complexity.
By demonstrating the difficulty of gray area moderation, we expose the shortcomings of current approaches and tools, while demonstrating the central role of experienced human moderators in maintaining healthy governance. Our findings have implications for addressing the tensions between moderators' labor conditions and community governance quality. We provide insights for integrating deliberation in moderator tool design. 

The rest of the paper is structured as follows. After contextualizing gray area cases in the literature of moderation practices and the sociotechnical systems in which they are embedded, we introduce the dataset of moderation logs and provide an encompassing description of the gray area, comparing and contrasting it with undisputed cases. Next, we address the crucial issue of automation in moderating the gray area by providing information-theoretical estimates of the difficulty of gray area cases and benchmarking state-of-the-art language models. Finally, we unpack the factors associated with LLMs' lackluster performance and discuss them in light of our novel understanding of the gray area.

\section{Related Work}\label{sec:rel_work}
\subsection{The Labor of Volunteer Moderation}


Platforms increasingly include volunteers in their moderation processes. For example, major social media platforms, including X and Facebook, have switched away from moderation based on fact-checkers to community-based moderation \citep{wojcik_birdwatch_2022}. Volunteer moderation represents a critical form of digital labor that sustains online communities at scale. \citeauthor{matias2019civic-4be}'s seminal work conceptualized volunteer moderation as ``civic labor,'' highlighting how unpaid moderators create and control public discourse for millions while their contributions remain largely invisible~\cite{matias2019civic-4be}. Recent quantitative analyses have demonstrated the economic value of this hidden labor. The value of volunteer moderation on Reddit alone is estimated to be a minimum of \$3.4 million per year, equivalent to 2.8\% of Reddit's 2019 revenue \cite{li2022measuring-0b5}. 

The nature of moderator work extends beyond simple rule enforcement. This labor encompasses not only content evaluation but also community norm development, conflict resolution, and platform governance~\cite{li2022all,roberts2019behind}. Volunteer moderators develop a sophisticated understanding of their community contexts, adapting global platform policies to local norms and values~\cite{seering2020metaphors-5f9,Gilbert2020,weld2021making-1bc}. 

Volunteer moderators face growing tensions. Their decisions are often contested \cite{koshy2023measuring-4aa,weld2024perceptions-03c}, they work under high expectations with little platform support \cite{schmitz2025from-b97}, and many experience burnout and turnover \cite{schpke-gonzalez2022why-2e8}. Our work highlights difficult moderation cases that may result from these challenges and points to ways to improve their working conditions.

\subsection{Collaboration in Moderation}

Effective moderation requires deliberation. Moderators possess valuable expertise about local norms and contexts that automated systems fail to capture~\cite{seering2019moderator-86d}. This expertise becomes crucial when adjudicating content that may appear problematic to outsiders but serves legitimate community functions \cite{10.1145/3613904.3642333}. 
\citeauthor{jhaver2019human-machine-005} demonstrated that seemingly clear community rules require contextual interpretation, with moderators applying different standards based on user history, intent, and community dynamics~\cite{jhaver2019human-machine-005}. Linguistic work further indicates that approved and removed comments can look strikingly similar~\cite{samory2021positive}. Such ambiguities are solved through collaboration and knowledge passing within moderation teams which are of increasing interest to the research community \cite{koshy2024venire-5f7,chen2023case-2fe}. Yet, we lack empirical insight into how disagreements arise within teams and how they are resolved---a gap this work begins to address. Specifically, our work characterizes the nature of disputed decisions in moderation teams, providing the empirical foundation needed to design transparency mechanisms that acknowledge the contestability in content moderation. Our work also reveals a significant part played by automated moderation systems in dispute cases. Next, we review work outlining the uncertainty introduced by automated systems in the moderation process. 

\subsection{Uncertainty in Automated Moderation}

Automated content moderation systems have become widely adopted by major platforms. For example, Meta reports that about 95\% of hate speech content removed on Facebook and Instagram is proactively detected by AI tools rather than via user reports.\footnote{\url{https://about.fb.com/news/2020/11/measuring-progress-combating-hate-speech/}} On Reddit, large language models (LLMs) are increasingly proposed to ease the workload of volunteer moderators. When prompted with subreddit rules, they can detect policy violations with high accuracy~\cite{kumar2024watch}. Especially, automated moderation systems for Reddit show promise to identify its variety of community-sensitive norms \cite{park2021detecting-605, he2023cpl-novid-a60,zhan2024slm-mod-1bd,goyal2025momoe-621}. 
However, while classifiers for harmful language demonstrate high performance on benchmark datasets, their effectiveness in practice for moderation tasks has been questioned~\cite{gillespie2018custodians,gorwa2020algorithmic-065}. 

Treating disputed cases the same way as straightforward ones could obscure weaknesses, as previous work warns against oversimplifying nuanced disagreement~\cite{pavlick2019inherent}. Annotation difficulty, operationalized as annotator disagreement, has also been identified as a confounding factor in human-AI alignment~\cite{alipour-etal-2025-robustness}. More fundamentally, a central tension lies in the construction of ground truth for training machine learning models. Much of the computational literature assumes a single authoritative label for training and evaluation, yet linguistic and cultural theory suggests that disagreement is often inherent rather than accidental \cite{Cabitza_Campagner_Basile_2023}. In this work, we focus on the political nature of labeling by looking at contested cases and show to what extent gray area content is more challenging for models. 

\section{Data}\label{sec:data}
Following the approach outlined in~\citet{samory2021positive}, we collect Reddit’s moderation logs, which are records of all moderation actions taken in a community. As part of the OpenModLog transparency initiative, subreddits can opt in by inviting the \textit{publicmodlogs} account, after which the bot republishes the community’s moderation logs to a public feed accessible to anyone. 

\subsubsection{ModLog Data Curation.}
From the raw moderation logs, we retain only actions corresponding to comment or post approvals and removals, which we grouped into two general categories (approve and remove). Other moderation actions, such as ``edit settings'' or ``sticky'' are ignored as they do not directly signal a sanctioning action taken at a comment level. To ensure robust statistical power, we selected the top 24 subreddits based on the volume of available moderation data. Specifically, we selected communities with sufficient activity to yield at least 250 disputed cases per outcome (approve/remove) alongside equivalent undisputed baselines. As these communities must opt-in to public logging to be included, our sample mainly consists of political and controversial subreddits (e.g., \textit{r/moderatepolitics}, \textit{r/conspiracy}, \textit{r/socialism}). While this selection is not representative of Reddit as a whole, this selection yields a dense dataset of \textbf{4.3 million moderation actions}. This full longitudinal dataset is used for the characterization in Section~\ref{sec:rq1}, while the balanced subset of 24,000 cases (described in \textit{Evaluation Subset}) is reserved for model benchmarking in Section~\ref{sec:rq2}. The full list is available in Appendix Figure \ref{fig:human_bot_mod_per_sub}.

To potentially separate human from automated moderation, we flag accounts with usernames containing substrings such as ``auto,'' ``bot,'' or ``modlogs.'' In total, 850 unique moderators were identified, of which 783 were classified as human moderators and 67 as automated moderators. In total, moderators took 4{,}272{,}178 actions
with humans taking 55.9\% and bots taking 44.1\% actions. Note that this number only points to the ``actions'' and not individual comments, which add up to 3.7M. Appendix Figure~\ref{fig:human_bot_mod_per_sub} reports bot and human moderator counts per subreddit. To avoid potential shifts in moderation behavior associated with LLMs, we discard actions logged after the release of ChatGPT on November 30, 2022. We refer to the resulting dataset as ``ModLog.'' Figure~\ref{fig:modlog_actions_timeframe} in the Appendix shows the temporal distribution of actions.

\subsubsection{Gray Area ModLogs.} Figure~\ref{fig:worked_example}.a shows an example of multiple moderation actions made on a single comment. First, the AutoModerator removes the comment based on a word filter. This decision is reversed by a human moderator (human mod 1), which is further contradicted by another human moderator (human mod 2). The comment eventually gets removed; however, these kinds of contrasting moderation actions exemplify what we call a ``gray area.'' Simply put, the comments that are in the gray area in our study have two or more contrasting moderation decisions made by two or more unique moderators. A comment and all of its associated decisions together constitute a case.

\paragraph{Case Stratification.}
For each case, we retain the full sequence of actions and moderators, along with considering the \emph{last} action as the ground truth on the basis that it represents the final decision that persisted on the platform. We categorize each case into one of four mutually exclusive strata based on whether there is within-case disagreement (multiple distinct actions and moderators) and there are any bot participants (see Figure~\ref{fig:worked_example}.b):

\begin{itemize}
    \item \textbf{gray-human}: more than one unique moderator and more than one unique action, and all moderators involved are human (i.e., disagreement among human mods).
    \item \textbf{gray-bot}: more than one unique moderator and more than one unique action, and at least one bot among the moderators.
    \item \textbf{undisputed-human}: not gray, and with only human moderator(s) involved. 
    \item \textbf{undisputed-bot}: not gray, and at least one bot is present. 
\end{itemize}

\subsubsection{Moderator Experience.}
Conflict in moderation actions can come from a variety of factors, including the expertise of the human moderators. To study how moderator experience affects gray area dynamics, we augment the dataset with moderator experience at the time of action, calculated as the number of days between the action and the moderator's joining date. The data for joining dates is collected from multiple sources such as the subreddit front page, moderation page, and wiki from the wayback machine, as well as other lists curated by Reddit users.\footnote{\url{https://www.reddit.com/mod/interestingasfuck/moderators/}} Within the ModLog, we identify 783 unique human moderators, of which 465 (60\%) overlap with the join-date dataset. Their join dates as moderators span March 2012 through July 2025. Appendix Figures~\ref{fig:mod_meta_joined_dist} and \ref{fig:mods_in_meta_per_sub} report the distribution of join dates and the per-subreddit counts of identified moderators.

\paragraph{Evaluation subset.}
One of the core contributions of this study is to juxtapose disputes in practical moderation decisions with the LLMs' alignment with the final decision in the moderation chain. 
To evaluate language model alignment on both gray area and undisputed cases, we construct a balanced per-subreddit dataset, ``ModLog Sample.'' From each subreddit, we sample 1{,}000 instances: 250 gray area cases ending in \textit{remove}, 250 gray area cases ending in \textit{approve}, and an equivalent split for the undisputed cases. We use this evaluation dataset of 24K samples in Section \ref{sec:rq2} and Section \ref{sec:rq3}. Before measuring the LLM alignment on this subsampled dataset,  we leverage the entirety of ModLogs to characterize the moderation rationale and content in the gray area. 

\section{Characterizing the Gray Area}\label{sec:rq1}
To characterize the gray area of moderation, we unpack moderation cases and explore them from multiple angles. Specifically, we examine the moderation actions that document the process of disagreement, the rationales that moderators may provide to justify their actions, and the contents of the moderated comments. 

\subsection{Sequences of Moderation Action}\label{sec:rq1:actions}
Gray area cases account for 13.54\% of all moderation actions (\(n=578{,}251/4{,}272{,}178\)). Within these disputed cases, 53.86\% involve only human moderators (\(n=311{,}465\)) and 46.14\% include at least one bot moderator (\(n=266{,}786\)). Analyzing the sequences of moderation actions that occur to adjudicate each case reveals two distinct patterns involving the correction of overmoderation by automated bots and the role of experienced moderators on final labeling decisions.

\paragraph{Bot actions are reversed by human mods.}
Restricting to gray areas and classifying moderation action sequences by the type of the first and last moderator, we observe that the overwhelming majority are $bot\!\to\!human$ at \textbf{87.00\%}, followed by $human\!\to\!human$ at 7.79\%, $bot\!\to\!bot$ at 4.47\%, and $human\!\to\!bot$ at 0.73\%. Because the gray area requires conflicting actions within a case and we treat the last action as ground truth, the dominance of $bot\!\to\!human$ sequences indicate that human moderators typically reverse or correct initial bot decisions.

This interpretation is reinforced when examining the distribution of first actions within disagreements. Among the first actions taken by bots, 95.49\% are removals and only 4.51\% are approvals. In contrast, human moderators exhibit a much more balanced pattern of 68.67\% approvals versus 31.33\% removals. The asymmetry between these two groups highlights a systematic tendency of bots to remove content preemptively, whereas humans are substantially more likely to reinstate it. By spelling out these dynamics, we can see that in \textbf{93.46\%} of all $bot\!\to\!human$ sequences the initial bot removal was directly followed by a human approval. We refer to this pattern as \textbf{overmoderation by bots} where automated systems disproportionately err on the side of removal, and human moderators subsequently intervene to restore content that was judged acceptable. For downstream analysis, we discard all $bot\!\to\!bot$ and $human\!\to\!bot$ action sequences, which are a large minority, and by manual inspection, appear as resulting from bot configuration issues. 

\paragraph{Experienced human moderators corrected inexperienced ones.} 
To investigate the role of tenure in conflict resolution, we focused on the subset of disputed cases involving human moderators. We isolated pairs of contradictory actions within these threads, strictly filtering for instances where two distinct human moderators acted on the same target. Analyzing the tenure difference across 11,587 such pairs shows that the intervening moderator typically has significantly more experience than the initial actor. On average, the moderator performing the reversal is 51 days more senior than the moderator being corrected (95\% CI: $[37, 65]$ days). This significant seniority gap indicates that the ``gray area'' of moderation is frequently resolved through a hierarchy of competence, where experts intervene to rectify decisions made by less experienced peers, thereby maintaining consistency in community governance.

\subsection{Moderation Rationales}\label{sec:rq1:rationales}

\begin{figure}[ht]
    \centering
    \includegraphics[width=\linewidth]{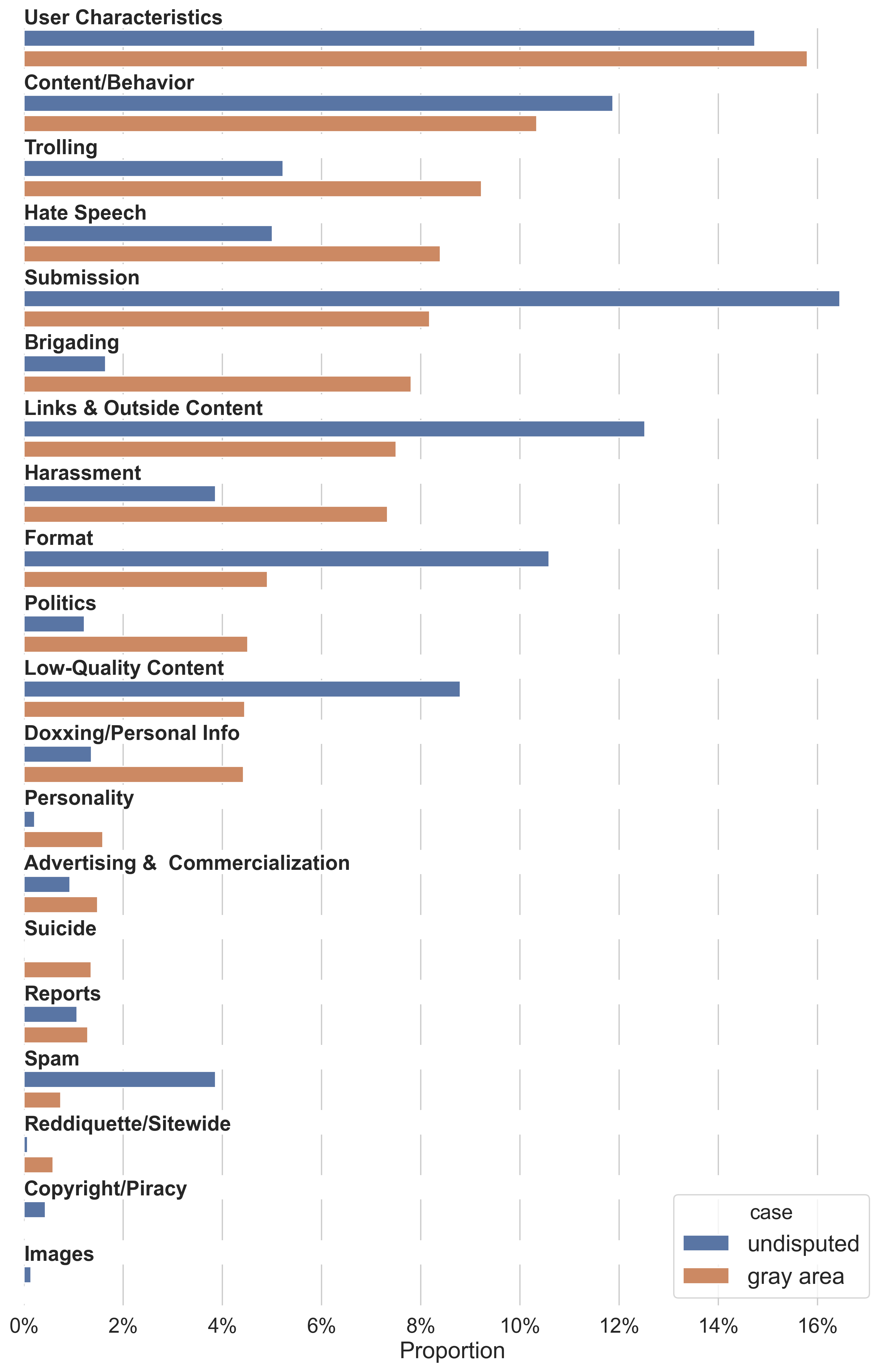}
    \caption{Share of moderation cases by stratum (gray vs. undisputed) across rule categories. Gray-area cases are relatively overrepresented in trolling, brigading, and doxxing, while spam, link-only, and formatting violations make up a larger share of undisputed cases.}
    \label{fig:details}
\end{figure}

Beyond the actions performed on a case, the ModLog dataset also records reasons provided by moderators, bots, or humans while approving or removing a comment. Such moderation rationales are free-text messages added either during the bot action or during the manual moderation process by humans. Out of the 3.7M comments across 24 subreddits in this study, only 594K had interpretable reasons provided in any of the associated moderation actions. Though somewhat underutilized as a feature of transparency in moderation, the reasons provided can still expose specific enforcement dimensions that lead to disputes. 

\subsubsection{Rationale categories.} We use the rule taxonomy provided by \citeauthor{fiesler2018reddit} to map free-text moderation rationales to rule categories, using a mixed-methods procedure. In total, we found 29K unique moderation rationale texts for 594K comments mentioned above. We group 29K rationales into 472 clusters based on very high ($>0.9$) cosine similarity using word count vectors \cite{phadke2024characterizing}. Each of those 472 clusters was further manually annotated with rule categories. For instance, rationales such as ``high report count'' or ``removed with 10+ reports'' were mapped to the ``Reports'' category. Similarly ``Possible hate speech''  or ``racial slur'' were mapped to the ``Hate speech'' category. We also added new categories, such as brigading or user-level removals (due to account age or low karma), that were not covered in the previous taxonomy \cite{fiesler2018reddit}. 
On Reddit, every comment is part of a larger conversation thread. We find that almost 40\% of the comments entered the ModLog due to their entire conversation thread being put under review. This illustrates how varied moderation actions are, with a substantial portion of moderation actions tied to the submissions themselves, without accounting for the comment-level content. 
To better capture the nuances of moderation rationales at the comment level, we remove all comments with submission-level reviews. For the ease of interpretation, we present broader results in Figure \ref{fig:details} and subreddit-level results in the Appendix Figure \ref{fig:sub_split_details}.

\subsubsection{Gray area rationales.}  

Figure \ref{fig:details} displays how comment moderation is spread across disputed and undisputed cases belonging to different categories. For example, almost 10\% of the gray area cases belong to the trolling category, while only 5\% of the undisputed ones are. In general, categories like hate speech (9\%), harassment (7\%), brigading\footnote{a group of users going from one subreddit to another to change or manipulate the discussions} (8\%), and suicide (1.5\% in disputed, with almost no undisputed cases) make for larger proportions of disputed cases than the undisputed. This may suggest that disputes in moderation may come from content that requires more context or subjective interpretation. A closer manual inspection reveals that many initial automated moderation actions are based on keywords, which are then reversed upon review. For example, a comment saying \textit{``fellow europoor here. i think i’ll keep my body fat percentage up until spring''} was initially removed by AutoModerator for the use of the word ``europoor'' caught in the racism keyword filter. Automated moderation actions based on simplistic keywords, later disputed by human moderators, may point to functional inefficiencies in the moderation process. In subreddits related to politics, categories like hate speech make for a lesser proportion of disputes ($15\%$) (Appendix Figure \ref{fig:sub_split_details}). Interestingly, manual inspection still pointed to blunt hate keyword filters such as ``idiot'' or ``insane'' which are used by AutoModerator to remove comments like \textit{``I'm scrolling comments like an idiot...''}. However, these actions are rarely disputed by human moderators, pointing to different cultures of content curation and due process across different subreddits.

\begin{table*}[ht]
\centering\sffamily\small
\resizebox{0.9\textwidth}{!}{%
\begin{tabular}{@{}p{1.5cm}p{3cm}p{8cm}p{5cm}@{}}
\toprule
\textbf{Category}  & \textbf{Topic examples} &\textbf{Comment example} & \textbf{Explanation}                                                      \\ \midrule

\textit{Non constructive}   & {6: rip, beautiful, lmaooo, lib; 49: based, ok, edited, snapchat
}                          & Rip consumeProduct &Users commenting with surface-level reactions or low-effort texts
\\
\textit{User complaints}    & {0: ban, rules, rule, mods;
4: idiots, cesspool, stupid, site
}                          & Why not just explain that then? They were deleting comments that didn't dox him aswell. I'm pretty sure a simple "guys stop posting this video is considered doxxing when you give out his name, and the admins are threatening to ban us".  & Users criticizing moderation decisions, policies, and tensions with Reddit's governance                                                                                                            \\
\textit{User interventions} & {26: concert, mod, relax, jannies;
28: behave, hilarious, tho, yall} & Y’all’s can’t behave                                                                                                                                                                                                                                                                                                                                                                                             & Users intervening in discussions to call out or de-escalate heated conversations.                                                                                                                  \\
\textit{Bots}               & {42: score, submissions, scores, confidence;
18: bloop, bleep, threads, linked} & We are spam filtering all comments by default in response to increased aggressiveness and unpredictability of “Anti-Evil Operations”I am a bot, and this action was performed automatically. Please contact the moderators of this subreddit if you have any questions or concerns.                                                                                                                              & Content produced by bots, including actions performed automatically and in bulk, user-created automations, typically to track content, scores, and censorship. \\
\textit{Moderation actions} & {70: evade, removals, pretext, variations;
69: violating, content, nz, copyright}& {[} Removed by Reddit {]} {[} Removed by reddit on account of violating the {[}content policy{]}(/help/contentpolicy). {]}                                                                                                                                                                                                                                                                                       & Moderation actions that alter comments or reply to them to provide explanations.                                                                                                                            \\
\textit{Borderline content} & {92: ironic, sus, penis, cope;
17: cringe, npc, heroes, epic}                                  & Keanu moment                                                                                                                                                                                                                                                                                                                                                                                                     & Low effort content, spam, attacks, and content that does not constructively advance the discussion.                                                                                         \\
\textit{General topics}     & 1: {bernie, candidate, gun, voting;
11: house, exclusive, summit, treasury}                       & House Democrats say Facebook, Amazon, Alphabet, Apple enjoy ‘monopoly power’ and recommend big changes                                                                                                                                                                                                                                                                                                           & Discussions about current events, politics, and finance.                                                                                                                                           \\
\textit{Subreddit topics}   & {10: iran, iranian, saudi, iranians;
35: collapse, suffering, climate, sustainable}& The Message Of The Anunnaki An Incredible Text First Published In   “We are already here, among you. Some of us have always been here, with you, yet apart from, watching, and occasionally guiding you whenever the opportunity arose.”                                                                                                                                                                         & Discussions about the specific topic of a subreddit.                                                                                                                                               \\ 
\bottomrule
\end{tabular}%
}
\caption{Overview of the thematic categories of gray-area cases, identified via topic modeling and iterative coding.}
\label{tab:topics}
\end{table*}

\subsubsection{Undisputed rationales.}
While categories like hate speech and harassment see a mixed proportion of gray area across subreddits, some rationales related to the use of external links or format are largely undisputed. This is not surprising given that format requirements such as comment length or title character limit have straightforward interpretations. For example, almost 11\% of the undisputed cases fall in the format category as displayed in Figure \ref{fig:details}. While most subreddits have fewer disputes in comments reviewed for Content (12\% of the undisputed cases), gaming subreddits like \textit{r/KotakuInAction} see a high proportion of disagreements ($61\%$ in Figure \ref{fig:sub_split_details}), again, likely coming from ineffective keyword filters. For example, a comment with text \textit{``...because they were already too entrenched in education...''} was removed by AutoModerator with the rationale \textit{``Transgender Topic ban; Matched: 'tren'''}. In contrast, Spam, though largely detected based on keywords, sees fewer disputes. This may indicate that most subreddits have relatively accurate spam filters based on the style, formatting, and links to known spam rings. 

Overall, analyzing moderation rationales across the ModLog reveals that a large share of actions reflect ambiguous or larger interventions---such as wholesale submission removals---rather than targeted enforcement of specific rule violations. These actions often provide little clarity on the exact rationale, highlighting both the limits of transparency and the difficulty of interpreting moderation at scale. Moreover, this analysis draws only from the subset of ModLogs where some rationale is provided. To more fully capture the dynamics of moderation disputes, we next turn to analyzing the comment content itself, as described below.

\subsection{Moderated Content}\label{sec:rq1:topics}

To gain an encompassing view of the gray area at scale, we perform topic modeling of the content that is subject to dispute. We adapt the BERTTopic \cite{grootendorst2022bertopic} approach to identify cross-cutting issues across subreddits. To do so, we need to remove the topical structure of subreddits from the text of gray area comments, which we achieve through an adversarial neural network architecture that leverages the adversarial debiasing framework of \cite{zhang2018mitigating} to strip away the subreddit bias, while employing the contrastive learning principles of \cite{oord2018representation} to ensure the resulting embeddings remain topically coherent (we defer the methodological details for this procedure to the Appendix \ref{app:debiasing}). We then extract topics on the debiased embeddings with BERTTopic, setting the HDBSCAN algorithm \cite{mcinnes2017hdbscan} with a minimum cluster size of 50 and a leaf-based cluster selection method. We improve topic representations by removing stopwords that are too frequent or infrequent ($max\_df=0.4$, $min\_df=0.05$), using a CTFIDF model set to reduce the importance of frequent words through BM25 weighting, and distinguishing words in the topics using a MaximalMarginalRelevance model with a diversity of 0.5. This procedure yields 116 topics. Then, we summarize the topics through an iterative coding procedure. We analyze the words and comments most associated with each topic, and iteratively refine a set of open codes, which we consolidate into seven overarching thematic categories (exemplified in Table \ref{tab:topics}). 

\paragraph{{Non-constructive comments}} Three of the seven themes are attributable to non-constructive comments, broadly defined. \textbf{Borderline content}---including low-effort posts, advertisements, and personal attacks---shows the necessity to balance disruptive behavior with promoting participation. \textbf{General discussions} of politics, economics, and current events include hot-button issues that may be difficult to moderate. \textbf{Subreddit-specific} cases, by contrast, reveal how localized norms condition governance. 

\paragraph{{User complaints.}}
A prominent category centers on user complaints directed at policies, moderators, and platform governance. These complaints manifest users' distrust of the moderation system, as individuals question not only the specific reasons behind removals but also the broader legitimacy of enforcement, which is depicted as opaque, inconsistent, or overly punitive. The intentions of moderators themselves and the state of the platform are often put into question, especially in the context of alleged shadowbanning incidents.

\paragraph{{User interventions.}}
Alongside criticism, a second pattern emerges in which users attempt to regulate discourse themselves by de-escalating conflicts or calling attention to inappropriate behavior. Users may themselves censor others or self-censor. These interventions may be read as a peer governance where ordinary users uphold the community norms, while simultaneously blurring the boundaries between user and moderator roles. 

\paragraph{{Bots.}}
Another theme involves content generated by bots that relay moderation decisions. This includes publicly communicating automated moderation decisions or human moderators locking threads and removing content in bulk. In several cases, the latter was publicly justified by moderators as preventive actions to dispel the intervention of ``Anti-Evil Operations,'' an account operated by Reddit's corporate moderators which can lead, in extreme cases, to shutting down subreddits and replacing moderators. Besides moderation automations, user-deployed bots appear frequently, for example, through comments that signal cross-posts. The scalability of bots, therefore, represents an infrastructural layer that both alleviates and worsens moderator workload. 

\paragraph{{Moderation actions.}}
Explicit moderation interventions, such as removals and removal explanations, constitute another category. These visible signals make enforcement legible to community members, serving as both a sanction and a pedagogical tool. However, their presence in the gray area raises questions about the archiving of moderation cases.

Taken together, these categories reveal the complexity of gray-area moderation. Gray area moderation may not simply be a matter of error correction but a constitutive practice that shapes the conditions of online discourse itself. Next, we investigate the extent to which computational models succeed or fail in adjudicating gray area cases.

\section{The Difficulty of Gray Area Cases}\label{sec:rq2}
Gray area cases provide a natural setting to test automated moderation. They can demonstrate both the potential of machine learning to support moderators at scale and help estimate the limits of automated means in making complex moderation decisions. Next, we estimate the difficulty of gray area cases using Pointwise $\mathcal{V}$-Information (PVI) and compare binary classification performance across 6 LLMs. We used the subsampled set of 24 balanced moderation logs described in Section \ref{sec:data} for this analysis. 

\begin{figure}[htb]
  \centering
  \includegraphics[width=.9\linewidth]{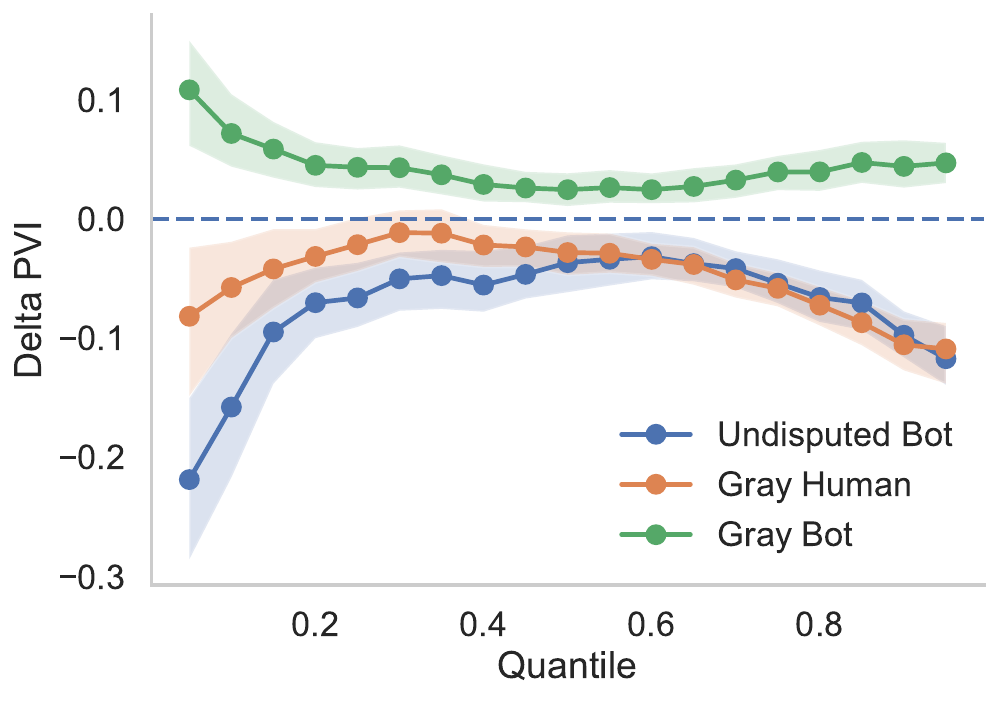}
  \caption{Differences in text difficulty (PVI) relative to the ``undisputed human'' baseline, with positive values indicating clearer (easier) text and negative values indicating more ambiguous (harder) text across quantiles. Shaded areas represent pointwise 95\% bootstrap confidence intervals.}
  \label{fig:shift}
\end{figure}

\subsection{Estimating the Difficulty of the Gray Area}\label{sec:rq2:pvi}

Pointwise $\mathcal{V}$-Information (PVI) 
is a measure of how much a model family $\mathcal{V}$ (e.g., BERT) can reduce the uncertainty about the true label when given the input, relative to a baseline that does not use the input~\cite{xu2020theory}. Following prior work~\cite{ethayarajh2022understanding, sen2023people}, for instance $(x_i,y_i)$ we define:
\begin{equation} 
\mathrm{PVI}_i
\;=\;
\log_2 p_{\theta}(y_i \mid x_i)\;-\;\log_2 \hat{p}(y_i).
\end{equation}
Here, $x_i$ is the comment text, $y_i \in \{approve, remove\}$ is the gold label, $p_{\theta}(y\mid x)$ is the model's predicted probability of the label given the text, and $\hat{p}(y)$ is the baseline that ignores the input. Intuitively, the first term rewards instances whose text makes the correct label highly probable, and the second term subtracts what a model could achieve without seeing the text at all. Higher PVI values indicate easier instances and negative values indicate hard or misleading instances. We defer methodological details of the finetuning and cross-validation procedure to Appendix~\ref{app:difficulty}. We also provide example comments with their respective PVI values across strata in Appendix Table \ref{tab:pvi_examples_gray} and in Table S1 of the supporting information (SI).

To compare text-based difficulty across case strata (e.g. gray-human), we examine how the distribution of PVI varies by stratum. Let $\mathcal{S}$ denote the set of strata and let $s^\star$ be the reference stratum (i.e. undisputed-human). For each stratum $s\in\mathcal{S}$ and quantile levels $\mathcal{T}\subset (0,1)$ we define $Q_s(\tau)$ a sample quantile of PVI in stratum $s$ at level $\tau$ where $\tau\in\mathcal{T}$. Then, for each stratum $s$, we define the \emph{shift function} relative to the reference stratum $s^\star$ as:
\begin{equation}
  \Delta_s(\tau)\;=\;Q_s(\tau)\;-\;Q_{s^\star}(\tau)  
\end{equation}
which is expressed in PVI units. Positive values indicate that, at percentile $\tau$, PVI in $s$ is higher (easier) than in the reference; negative values indicate lower (harder). We quantify pointwise uncertainty with a percentile bootstrap at each $\tau$, independently resample (with replacement) within stratum $s$ and within reference, recompute the quantiles and their difference to form the 95\% confidence band. Figure~\ref{fig:shift} shows the quantile level on the $x$-axis against the difference in PVI at that quantile in $y$-axis. For example, $\tau=0.9$ compares the 90th percentile in stratum $s$ to the 90th percentile in the reference stratum, which is undisputed-human.

Moderation cases that fall into gray-human and undisputed-bot are below zero across almost all $\tau$ and have consistently lower PVI relative to undispute-human. The gap is largest in the lower tail ($\tau\approx$ 0.05--0.2), shrinks in the median, then widens again towards the upper tail ($\tau \gtrsim 0.6$). Based on the patterns, moderation instances in these two strata, on the text alone, are harder to predict. On the other hand, in gray-bot, the curve is positive across $\tau$, with the largest gap in the lower tail and modest advantages around the high quantiles. This indicates that cases where human moderators reverse automoderator actions typically contain clearer text evidence for the final outcome than undisputed-human cases.  

\subsection{LLM Moderation of Gray Area Cases}\label{sec:rq2:llm}
To assess how well language models align with the final actions recorded in the ModLog, we evaluate four open-source models (\emph{Llama 8B, Qwen 7B, Llama 70B, Qwen 32B}) and two proprietary models (\emph{Gemini Flash and GPT-5 Mini}). Given a subreddit's name, its description, the set of community rules, and the comment under review, the model is asked whether the comment should be approved or removed. Except where the API lacks controls (e.g., \emph{GPT-5 Mini}), we use deterministic decoding ($temperature=0$, $top\_p=1$). We report macro-F1 over the binary labels and 95\% bootstrapped confidence intervals. Full checkpoint details and additional prompt design, parsing rules, and template-level analyses are provided in Appendix~\ref{app:difficulty}.

\paragraph{Per-stratum performance.}

\begin{figure}[htbp]
  \centering
  \includegraphics[width=0.95\linewidth]{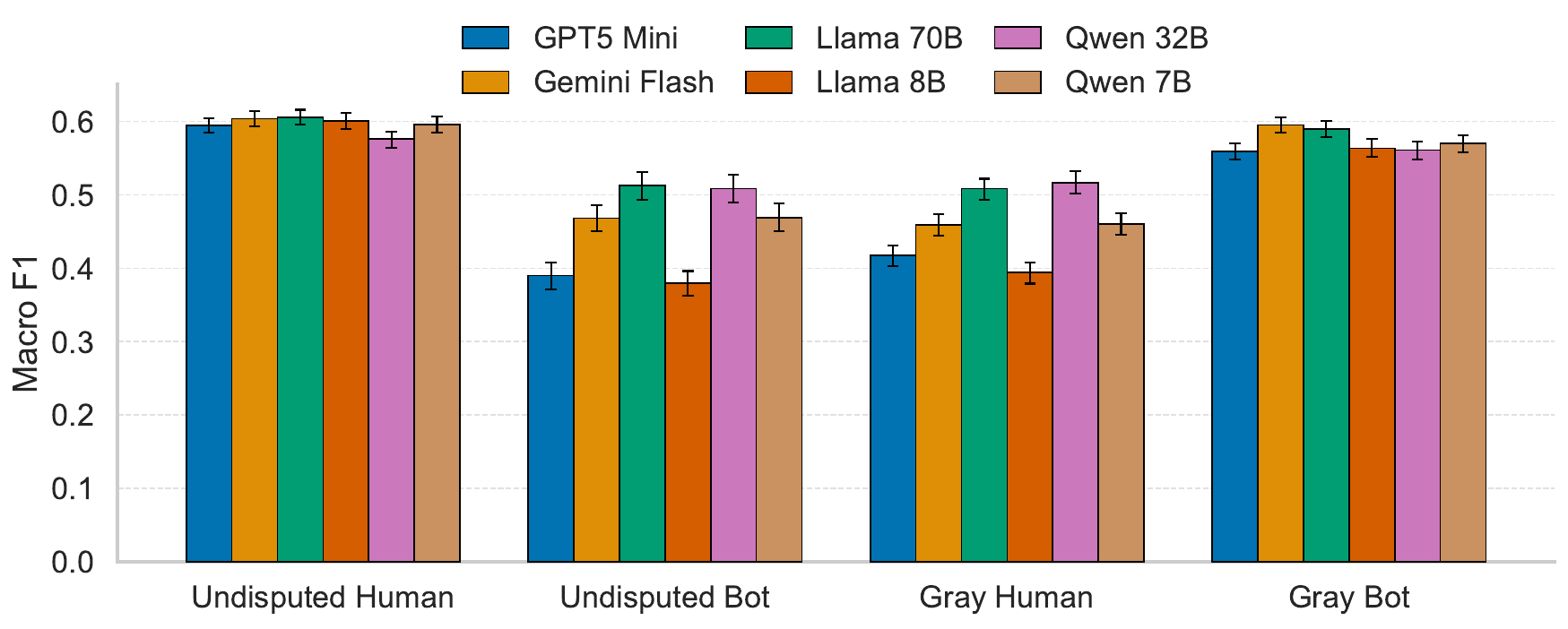}
  \caption{Macro-F1 and 95\% confidence intervals for different models across case strata.}
  \label{fig:llm_f1}
\end{figure}

Echoing our findings on instance difficulty in Section~\ref{sec:rq2:pvi}, the models exhibit two distinct performance tiers across the strata (Figure~\ref{fig:llm_f1}, Appendix Table~\ref{tab:f1-buckets}). The first tier comprises undisputed-human and gray-bot, where models achieve their highest macro-F1 scores. While undisputed-human nominally leads (e.g., \emph{Llama~70B}: $0.61$), the gray-bot stratum follows closely, performing nearly on par (e.g., \emph{Llama~70B}: $0.59$). This high performance correlates with the observed positive PVI curves, suggesting that instances where humans reverse automoderator actions often contain clear textual evidence. Conversely, the gray-human and undisputed-bot strata form a lower-performing tier. Scores here are consistently lower and tightly clustered (e.g., \emph{Llama~70B}: $\approx 0.51$ for both), reinforcing the observation that these instances are intrinsically harder to predict based on text alone.

\section{Gray Area Unpredictability Factors}\label{sec:rq3}
To characterize how observable features of a moderation case---such as text difficulty and moderator experience---affect the probability that the LLM's predicted moderation decision matches the individual moderator's action, we fit a Bayesian binomial mixed-effects model. Our analysis centers on \emph{Llama 70B} ($n=13,271$) predictions which is the best-performing model in our benchmark. We consider only actions performed by human moderators in the gray-human and gray-bot strata. The former reflects disagreement among humans, while the latter captures cases where humans corrected bot actions. 

\paragraph{Model specification.}
Let $Y_{i,j}\in\{0,1\}$ indicate whether the model's prediction matches the specific action taken by human moderator $j$ for case $i$. To account for where moderation practices vary by community, we include a random intercept for the subreddit $g[i]$. The model is specified as follows:
\begin{equation}
\begin{aligned}
\operatorname{logit}\,\Pr(Y_{i,j}=1) &= \beta_0
    + \beta_1\,\mathrm{PVI}_i \\ 
    & + \beta_2\,\mathrm{experience}_{i,j} \\
    & + \beta_3\,\mathrm{action_{i,j}} \\
& + \sum_k \gamma_k\,\mathrm{topic}_{k,i}
  + b_{g[i]}\,.
\end{aligned}
\end{equation}
The model estimates the baseline log-odds of alignment ($\beta_0$) alongside a subreddit-level random effect ($b_g$) drawn from a normal distribution $\mathcal{N}(0,\sigma^2_{\text{subreddit}})$. We control for the decision made by the human moderator using the binary variable $action_{i,j}$, which distinguishes between remove (i.e. reference category) and approve decisions taken by human moderators. 
We standardize continuous predictors---clarity of the text (PVI) and moderator's experience---by centering them and dividing their values by two standard deviations to make their scales comparable and interpretable. Table~\ref{tab:rq3_glmm} summarizes the factors driving alignment between \textit{Llama 70B} and human moderators by reporting the posterior means and 95\% Credible Intervals (CrIs) of the log-odds coefficients and, separately, the subreddit-level heterogeneity captured by the random-intercept SD.

\paragraph{Features that make alignment more (or less) likely.} Textual difficulty, measured by the PVI score, is the strongest positive predictor of alignment ($\beta_1 = 0.50$). As the text becomes clearer and less ambiguous (higher PVI), the likelihood that the model matches the moderator's decision increases significantly. Moderator experience also exhibits a positive association with alignment ($\beta_2 = 0.12$), where we observe higher agreement for decisions made by more experienced moderators. In addition, alignment is significantly higher for removal decisions than for approvals ($\beta_3 = 0.16$). 
Among content topics, subreddit-specific ($\gamma = 0.30$) and general topics ($\gamma = 0.23$) show strong positive effect with model alignment, potentially because the model leverages the provided community description and rules as context. 
We also observe positive effect in user complaints topic ($\gamma = 0.16$) which is a category defined by users criticizing moderation policies or platform governance.
On the other hand, other topic indicators have their credible intervals overlapping zero and are harder for the model to adjudicate in a consistent manner. One of such topics is the borderline content which shows no significant alignment effect, confirming that gray area trolling and subtle behavioral violations remain a blind spot for language models. 

\begin{table}[ht]
\centering
\sffamily
\small
\begin{tabular}{l r l}
\toprule
D.V: \texttt{alignment} & Post. Mean & 95\% CrI \\
\midrule
intercept & \textbf{0.11} & [0.08, 0.15] \\
PVI & \textbf{0.50} & [0.42, 0.57] \\
experience & \textbf{0.12} & [0.05, 0.19] \\
action=approve & \textbf{-0.16} & [-0.20, -0.12] \\
topic=subreddit & \textbf{0.30} & [0.22, 0.38] \\
topic=general & \textbf{0.23} & [0.16, 0.30] \\
topic=user complaints & \textbf{0.16} & [0.07, 0.25] \\
topic=mod actions & 0.12 & [-0.15, 0.38] \\
topic=borderline/low-effort & 0.05 & [-0.01, 0.11] \\
topic=bots & -0.03 & [-0.21, 0.16] \\
topic=user initiative & 0.07 & [-0.22, 0.36] \\
\midrule
\multicolumn{3}{r}{$\sigma_{\text{subreddit}}$ = 0.28\ [0.21, 0.38]} \\
\bottomrule
\end{tabular}
\caption{Posterior means and 95\% credible intervals predicting LLM alignment with human moderator actions. The reference category for action is \emph{remove}. Continuous predictors (difficulty, experience) are standardized. Subreddit random intercept captures between-subreddit heterogeneity. Boldface denotes 95\% CrIs that exclude zero.}
\label{tab:rq3_glmm}
\end{table}

\paragraph{Between-subreddit heterogeneity.}
We quantify the variability in moderation culture across different communities. The posterior mean of the subreddit standard deviation (SD) suggests that a typical community shifts the baseline log-odds by roughly $\pm0.28$. Starting from the baseline intercept \(\beta_0 = 0.11\), this yields
\(\operatorname{logit}^{-1}(0.11 \pm 0.28) \in [0.46, 0.60]\), which indicates that gray area unpredictability is driven as much by local community norms as it is by the inherent complexity of the cases themselves. We discuss these implications below.

\section{Discussion and Implications}\label{sec:discussion}
Taken together, our results paint a detailed picture of the moderation gray area and its characteristics. Far from being rare, the gray area affects one in seven moderation cases. Within it, we see four main types of cases discussed below.

\paragraph{Gray area as contested judgment.}
Difficult cases put moderators’ judgment to the test. Such cases often involve behaviors such as trolling or brigading, where intent is ambiguous, or severe situations such as doxxing and suicidal ideation, where interpretations are highly contextual. What counts as contentious also varies across communities, substantiating prior work on cultural differences in moderation \cite{chandrasekharan2018internet}. This suggests that subreddits develop their own moderation practices and due process, with implications for building transparent governance tools and for supporting community members through civic awareness and education.

\paragraph{Gray area as inefficiency.}
Routine corrections are also frequent---in fact, more so than difficult calls. Bulk actions, like reinstating a discussion thread after temporarily removing its comments, are arguably straightforward decisions that make up 40\% of the moderation log (limited to comment approval and removal actions). Along with these, we find that automated moderation bots produce 87\% of the gray area, vis-à-vis 44\% of total moderation actions. This is concerning because bots tend to over-moderate, raising fairness issues when there are too few human moderators to correct them. In fact, we find that automated moderation actions based on simplistic keywords often require correction, which are provided inconsistently across communities. More broadly, the volume of routine corrections, caused in large part by current automation tools available to moderators, points to functional inefficiencies in the moderation process, a finding which echoes Redditors'---and specifically, moderators'---demands of improved technical resources in recent protests against the platform~\cite{schmitz2025from-b97}. Gray area cases exemplify the misalignment between Reddit's moderation API, limited to transactional and atomic actions like comment approval and removal, and the need for complex workflows and deliberation processes that moderators engage in. 

\paragraph{Gray area as ambiguity.}
How to support moderators in such cases remains an open question. We add to recent work that shows the limits of LLMs as a primary avenue for automated moderation~\cite{zhan2024slm-mod-1bd}, and demonstrate that LLM performance is lackluster on gray area cases, which are akin to adversarial examples by construction. In particular, we find that cases that require deliberation among human moderators are harder to adjudicate through LLMs. We corroborate these findings through information-theoretic measures, which prove the intrinsic difficulty of such cases and, therefore, suggest that improvements in model capabilities may remain insufficient. 
Instead, sociotechnical designs that route ambiguous cases to multiple moderators~\cite{koshy2024venire-5f7}, or to more experienced moderators as seen in current practice, may be more promising.

\paragraph{Gray area as participatory governance.}
Our results suggest that moderator expertise plays a key role in tackling gray area cases and determining the final label for a comment. In this sense, contested cases could be read as a site of power exertion, but also as one of norm-setting, helping onboard less experienced moderators, identifying precedents, and converging on enforcement standards in moderation teams. Literature provides qualitative accounts of these practices \cite{Gilbert2020,doi:10.1177/14614448231190901}, which we complement with large-scale empirical findings. Yet, we find that the gray area cases involve several stakeholders in the governance process beyond the moderation team---not least, the authors of the moderated content, who may appeal moderator decisions, as well as the user community at large, contesting policies and taking an active part in norm enforcement. In this light, gray area cases offer an opportunity for participatory governance, embodying the social conflict that underpins all progress. Making the work of handling contested cases visible and transparent can improve online governance and open it to participation. Supporting such participation is likely to strengthen perceptions of fairness and community health \cite{atreja2024appealmod-20d,vaccaro2021contestability-93d,jhaver2019did-151,weld2024perceptions-03c}.

\paragraph{Limitations}
Our bot identification and account-state rule exclusion rely on heuristics and may leave residual cases. Tenure estimates for moderators are based on public join-date traces, and rows with implausible post-action timestamps were excluded. Moderator logs also provide only a partial view of moderation work. As \citeauthor{li2022all} note, much of this labor is ``invisible,'' including activities like responding to user messages, coordinating decisions, and debating policies in private channels. Prior qualitative work highlights the centrality of these tasks but also the difficulty of accessing them. Our analysis, limited to public logs, cannot capture these practices and therefore underrepresents the full scope of moderation. Finally, our data comes from communities that opted into public moderation logging. While this follows a more ethical research practice, it may reflect different transparency norms than other Reddit communities.

\paragraph{Ethics Statement}
The gray area of moderation is inherently contested, and we therefore make explicit the ethical considerations shaping this research. We follow established guidance for research using online data \cite{fiesler2018participant, proferes2021studying, bruckman2002ethical}. Our analysis involves content flagged as problematic, including hate speech, harassment, and personal attacks. Moderation data may contain offensive or sensitive material, and communities may legitimately disagree on appropriate standards. Some content, such as doxxing, carries serious risks if published or re-identified, including harm to victims, retaliation against moderators, or persecution of offenders. To reduce these risks, we report only anonymized or aggregate results and do not redistribute raw data or usernames, accepting the limits this places on replicability. We concur with prior ethical statements about users' expectations of privacy and longitudinal ModLog collection \cite{samory2021positive}.

\section{Conclusions}
This work provides an empirical characterization of the gray area in content moderation by examining contested cases in which at least two distinct moderators have expressed contrary judgments. 
We quantify the pervasiveness of disputed cases at one-in-seven decisions. Our findings provide a topical analysis of difficult-to-moderate cases, involving behaviors where user intent remains ambiguous, such as trolling and brigading. We also identify routine corrections, exposing significant inefficiencies in the current moderation infrastructure. Using Pointwise $\mathcal{V}$-Information, we establish that gray area cases are inherently harder to adjudicate than undisputed ones, with state-of-the-art large language models achieving substantially lower performance on disputed cases, while moderator experience emerges as central to effective resolution. These findings suggest prioritizing the support for deliberative processes among human moderators rather than pursuing automated resolution of contested cases---an approach our results indicate is fundamentally limited.

\section*{Acknowledgments}
We thank Eshwar Chandrasekharan, Eric Gilbert, and David Jurgens for formative discussions that helped shape the research.

\bibliography{aaai2026}

\section*{Paper Checklist }

\begin{enumerate}

\item For most authors...

\begin{enumerate}

    \item  Would answering this research question advance science without violating social contracts, such as violating privacy norms, perpetuating unfair profiling, exacerbating the socio-economic divide, or implying disrespect to societies or cultures?

    \answerYes{Yes}

  \item Do your main claims in the abstract and introduction accurately reflect the paper's contributions and scope?

    \answerYes{Yes}

   \item Do you clarify how the proposed methodological approach is appropriate for the claims made? 

    \answerYes{Yes}

   \item Do you clarify what are possible artifacts in the data used, given population-specific distributions?

    \answerYes{Yes}

  \item Did you describe the limitations of your work?

    \answerYes{Yes}

  \item Did you discuss any potential negative societal impacts of your work?

    \answerYes{Yes}

      \item Did you discuss any potential misuse of your work?

    \answerYes{Yes}

    \item Did you describe steps taken to prevent or mitigate potential negative outcomes of the research, such as data and model documentation, data anonymization, responsible release, access control, and the reproducibility of findings?

    \answerYes{Yes}

  \item Have you read the ethics review guidelines and ensured that your paper conforms to them?

    \answerYes{Yes}

\end{enumerate}

\item Additionally, if your study involves hypotheses testing...

\begin{enumerate}

  \item Did you clearly state the assumptions underlying all theoretical results?

    \answerNA{NA}

  \item Have you provided justifications for all theoretical results?

   \answerNA{NA}

  \item Did you discuss competing hypotheses or theories that might challenge or complement your theoretical results?

    \answerNA{NA}

  \item Have you considered alternative mechanisms or explanations that might account for the same outcomes observed in your study?

    \answerNA{NA}

  \item Did you address potential biases or limitations in your theoretical framework?

    \answerNA{NA}

  \item Have you related your theoretical results to the existing literature in social science?

    \answerNA{NA}

  \item Did you discuss the implications of your theoretical results for policy, practice, or further research in the social science domain?

   \answerNA{NA}

\end{enumerate}
    
\item Additionally, if you are including theoretical proofs...
\begin{enumerate}
  \item Did you state the full set of assumptions of all theoretical results?
    \answerNA{NA}
	\item Did you include complete proofs of all theoretical results?
     \answerNA{NA}
\end{enumerate}

\item Additionally, if you ran machine learning experiments...

\begin{enumerate}

  \item Did you include the code, data, and instructions needed to reproduce the main experimental results (either in the supplemental material or as a URL)?

    \answerYes{Code is available at \url{https://github.com/shayanalipour/the-gray-area}. The data will not be released due to ethical considerations.}

  \item Did you specify all the training details (e.g., data splits, hyperparameters, how they were chosen)?

   \answerYes{Yes}

     \item Did you report error bars (e.g., with respect to the random seed after running experiments multiple times)?

    \answerYes{Yes}

	\item Did you include the total amount of compute and the type of resources used (e.g., type of GPUs, internal cluster, or cloud provider)?

    \answerYes{Yes}

     \item Do you justify how the proposed evaluation is sufficient and appropriate to the claims made? 

   \answerYes{Yes}

     \item Do you discuss what is ``the cost`` of misclassification and fault (in)tolerance?

    \answerYes{Yes}

\end{enumerate}

\item Additionally, if you are using existing assets (e.g., code, data, models) or curating/releasing new assets, \textbf{without compromising anonymity}...

\begin{enumerate}

  \item If your work uses existing assets, did you cite the creators?

    \answerYes{Yes}

  \item Did you mention the license of the assets?

    \answerNA{NA}

  \item Did you include any new assets in the supplemental material or as a URL?

    \answerNA{NA}

  \item Did you discuss whether and how consent was obtained from people whose data you're using/curating?

    \answerYes{Access to Reddit ModLogs is described in Sections \ref{sec:data} and \ref{sec:discussion}}

  \item Did you discuss whether the data you are using/curating contains personally identifiable information or offensive content?

    \answerYes{Yes, and we are not releasing this dataset for ethical considerations}

\item If you are curating or releasing new datasets, did you discuss how you intend to make your datasets FAIR (see \nocite{fair})?

\answerYes{We are not releasing this dataset for ethical considerations}

\item If you are curating or releasing new datasets, did you create a Datasheet for the Dataset (see \nocite{gebru2021datasheets})? 

\answerNA{NA}

\end{enumerate}

\item Additionally, if you used crowdsourcing or conducted research with human subjects, \textbf{without compromising anonymity}...
\begin{enumerate}
  \item Did you include the full text of instructions given to participants and screenshots?
    \answerNA{NA}
  \item Did you describe any potential participant risks, with mentions of Institutional Review Board (IRB) approvals?
    \answerNA{NA}
  \item Did you include the estimated hourly wage paid to participants and the total amount spent on participant compensation?
    \answerNA{NA}
   \item Did you discuss how data is stored, shared, and deidentified?
   \answerNA{NA}
\end{enumerate}

\end{enumerate}

\appendix

\begin{figure}
  \centering
  \includegraphics[width=.9\linewidth]{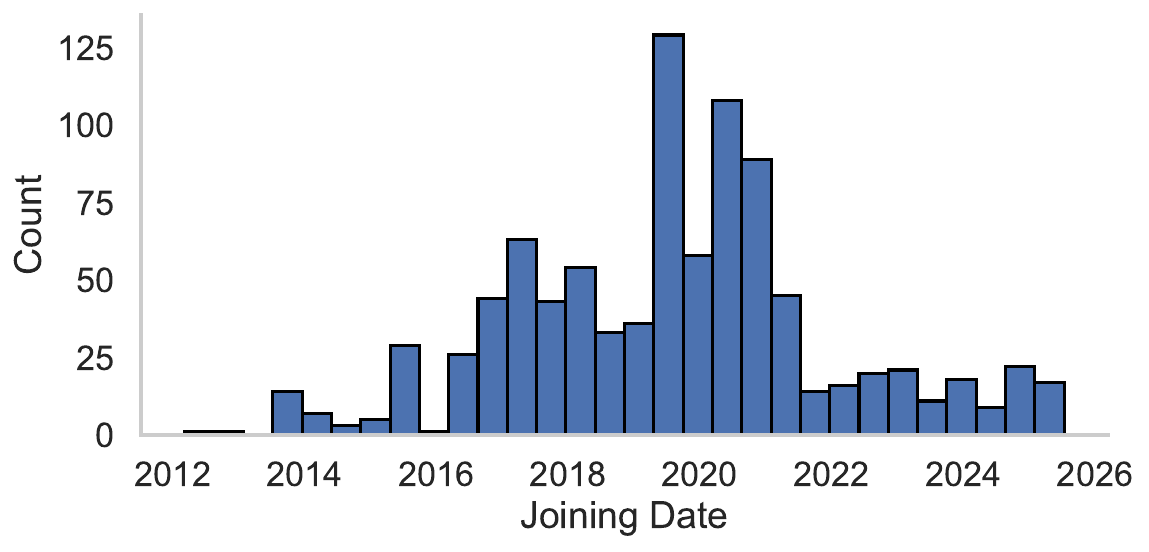}
  \caption{Distribution of moderators’ join dates in subreddits, used with action timestamps to calculate moderator experience (in days).}
  \label{fig:mod_meta_joined_dist}
\end{figure}

\begin{figure}
  \centering
  \includegraphics[width=.9\linewidth]{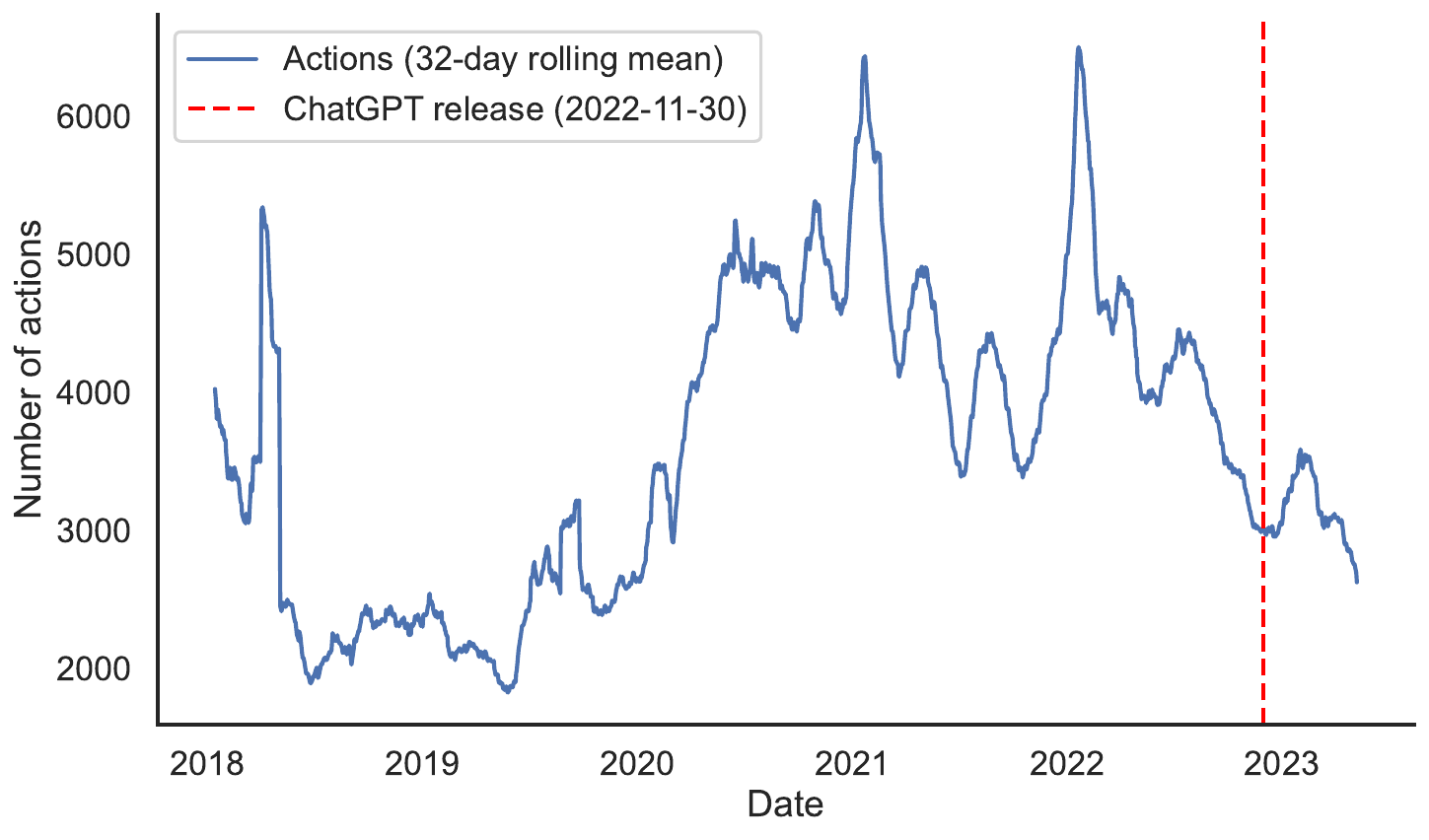}
  \caption{Smoothed (32-day centered rolling mean) daily count of actions in the ModLog dataset. The dashed vertical line marks the ChatGPT release on 2022-11-30 (cutoff).}
  \label{fig:modlog_actions_timeframe}
\end{figure}

\begin{figure*}
    \centering
    \includegraphics[width=\linewidth]{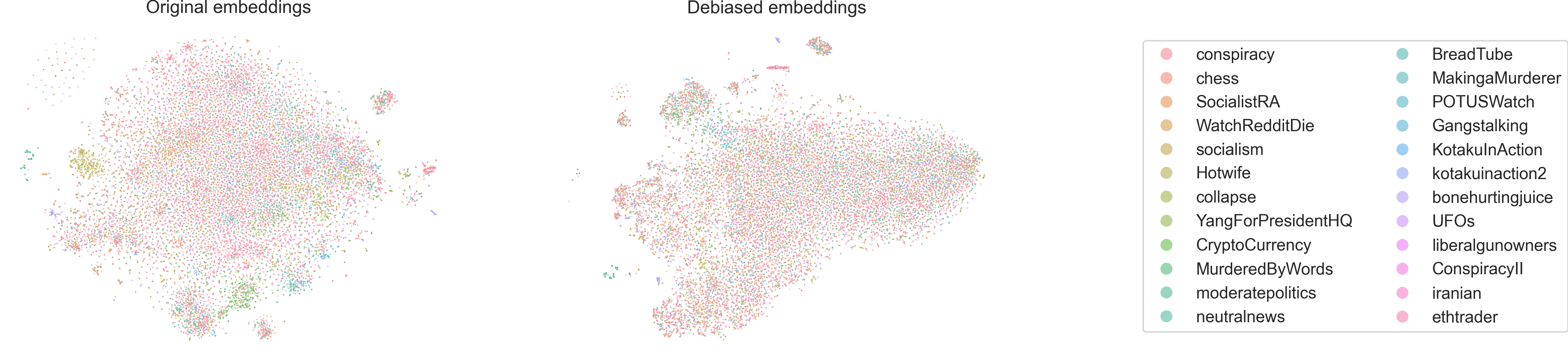}
    \caption{Sentence-embedding visualization of comments in gray-area before and after subreddit-debiasing. Each point is a comment, colored by its source subreddit. Left: original embeddings form tight, subreddit-specific clusters. Right: after training an adversarial network to remove subreddit identity, colors are mixed and community structure is minimized. We use these debiased representations with BERTTopic to surface cross-cutting issues that span subreddits.}
    \label{fig:debiasing}
\end{figure*}

\section{Debiasing Gray Area Content from Subreddit-Specific Topics}\label{app:debiasing}
We summarize the themes of gray area cases through BERTTopics, a technique that exploits the informativeness of contextualized document embeddings to identify similar clusters. However, since ModLogs pertain to various subreddits, the first-order differences between these communications are the topic and sociolect of the subreddits. To identify cross-cutting themes, we propose a novel method for generating a semantically meaningful, yet subreddit-agnostic, vector space. Our approach is designed to produce a low-dimensional embedding of comments that clusters by the core topic, rather than by the subreddit of origin. This methodology is particularly suited for analyzing gray area cases, where disputes manifest in the broader context of subreddit discourses.

\subsection{Model Architecture}
Our architecture consists of two interconnected neural networks trained in a competitive, adversarial framework. The first, a \textbf{Debiasing Head} ($D$), is a multi-layer perceptron (MLP) with a linear-ReLU-linear structure, mapping the high-dimensional initial embeddings ($x \in \mathbb{R}^{768}$), computed using \texttt{ConversationalBERT}, to a lower-dimensional, debiased representation ($z \in \mathbb{R}^{128}$). The second is a \textbf{Subreddit Classifier} ($C$), also an MLP, which takes the debiased embeddings as input and predicts the original subreddit label. This classifier serves as the adversary in our training regime.

\subsection{Training Regime}
The training process is an iterative optimization of both networks, with a loss function composed of four distinct components. Our goal is to train $D$ to minimize a composite loss while simultaneously maximizing the adversarial loss of $C$.

\begin{enumerate}
    \item \textbf{Topic Consistency Loss ($\mathcal{L}_{topic}$):} We first pre-cluster a subset of gray area documents using HDBSCAN on their initial embeddings to derive pseudo-topics. For a given mini-batch, this loss, formulated as a \textbf{gray area InfoNCE objective}, pulls the debiased embeddings of comments from the same pseudo-topic together while pushing them away from all other samples in the batch. This ensures that topic-level information is preserved during the debiasing process.

    \item \textbf{Gray Area Separation Loss ($\mathcal{L}_{gray}$):} This is a \textbf{margin-based contrastive loss} that minimizes the cosine similarity between the debiased embeddings of gray area and undisputed comments. It serves to create distinct, well-separated regions in the latent space for these two classes of documents, facilitating subsequent analysis.

    \item \textbf{Subreddit Debiasing Loss ($\mathcal{L}_{subreddit}$):} To remove subreddit-specific signals, we enforce an additional margin loss. This objective minimizes the cosine similarity between gray area comments from the same subreddit but belonging to different pseudo-topics. It forces the model to ignore subreddit-level linguistic conventions when they do not correspond to the core topic.

    \item \textbf{Adversarial Loss ($\mathcal{L}_{adv}$):} This is the central mechanism for disentanglement. The Subreddit Classifier ($C$) is trained to accurately predict a comment's subreddit from the debiased embedding $z$. Simultaneously, the Debiasing Head ($D$) is trained to maximize this loss, effectively learning to produce a representation that is unclassifiable by $C$. The loss for the Debiasing Head is thus the negative of the classifier's cross-entropy loss.
\end{enumerate}
The total objective function for the Debiasing Head ($D$) is a weighted sum of these components, optimized using the Adam optimizer:
\begin{equation}
    \mathcal{L}_D = \mathcal{L}_{topic} + \alpha \mathcal{L}_{gray} + \beta \mathcal{L}_{subreddit} + \lambda \mathcal{L}_{adv}
\end{equation}
\subsection{Evaluation}
We evaluate the efficacy of our debiasing approach by comparing the classification accuracy of two models: one trained to predict subreddits from the original embeddings (baseline) and another from the debiased embeddings. We observe a substantial decrease in accuracy from the baseline (from 0.36 to 0.11) towards random chance ($0.04 = \frac{1}{24}$), which demonstrates the model's success in removing subreddit-specific information from the vector space. 

This quantitative reduction in subreddit separability is qualitatively confirmed in Figure~\ref{fig:debiasing}. As shown in the visualization, the original embeddings (Left) form tight, subreddit-specific clusters, whereas after adversarial training (Right), the community structure is dissolved. The resulting colors are mixed, indicating that the debiased vector space is organized by semantic topic rather than the source subreddit.

\begin{figure}[ht]
    \centering
    \includegraphics[width=.99\linewidth]{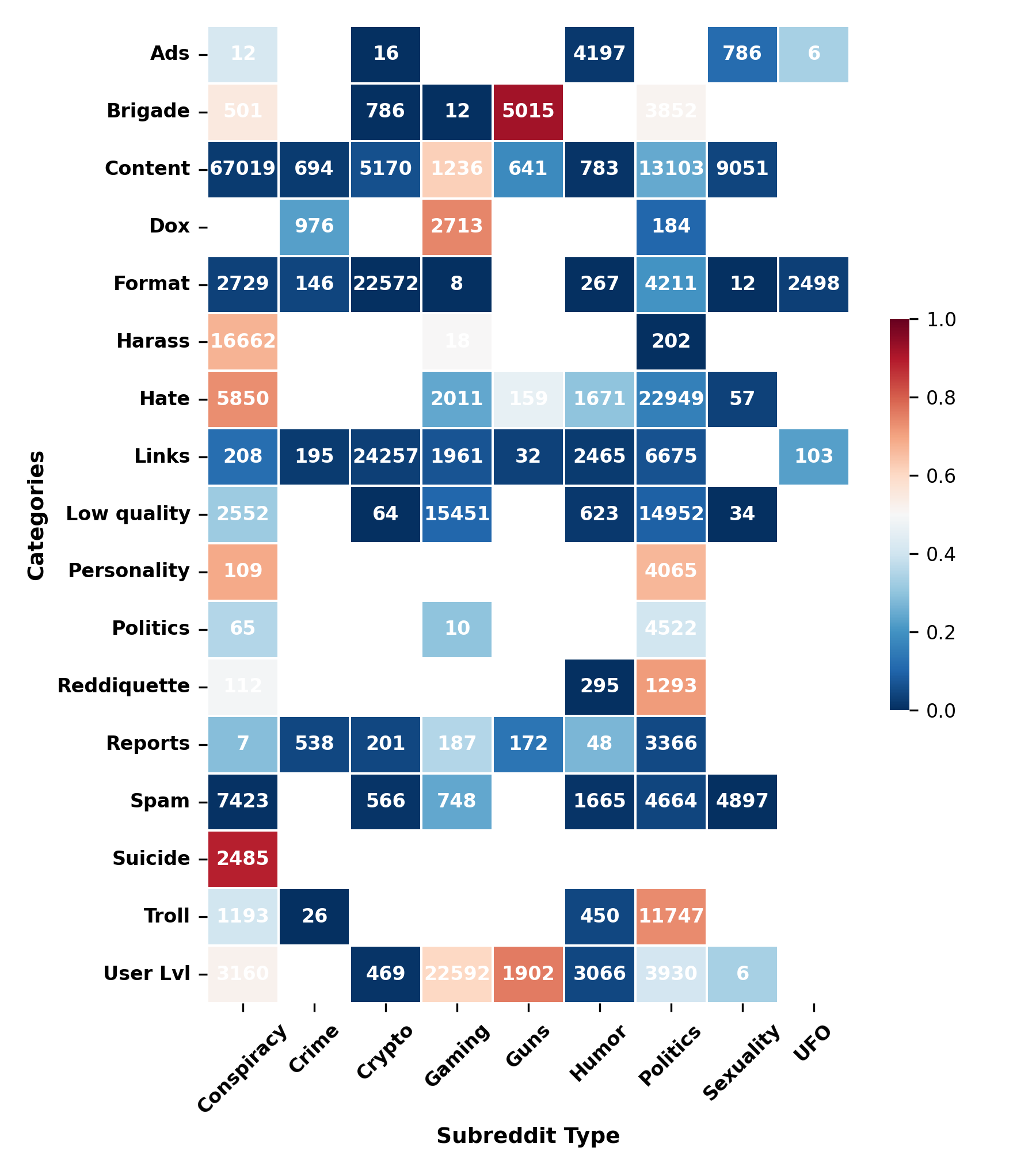}
    \caption{Undisputed (blue) and Gray (red) proportion of ModLog across different subreddit categories.}
    \label{fig:sub_split_details}
\end{figure}

\section{Difficulty of ModLog Dataset}\label{app:difficulty}
We study dataset “difficulty’’ from two complementary angles. First, we estimate instance-level text difficulty using pointwise $\mathcal{V}$-Information (PVI), which quantifies how informative a comment is for predicting the gold action. Second, we describe how we run and score LLM inferences across models and prompt templates to assess decision robustness.

\subsection{Estimating Text Difficulty}
PVI (defined in the main text) measures, for each comment, how much the input reduces label uncertainty compared to a label-only baseline. Higher values indicate “easier’’ instances for the model family; negative values indicate hard or misleading ones.

\paragraph{Finetuning procedure.}
To estimate $p_{\theta}(y\mid x)$ for PVI, we fine-tune a \texttt{RoBERTa-large} classifier with 5-fold stratified cross-validation where in each fold, the model trains on 80\% and predicts on the held-out 20\%. We use 3 epochs, AdamW (learning rate $2\times10^{-5}$, weight decay $0.01$, warmup ratio $0.06$), batch size 32, gradient clipping at 1.0, FP16, and the default \texttt{RoBERTa} tokenizer with dynamic padding. For each instance $i$, $\log_2 p_{\theta}(y_i \mid x_i)$ comes from the fold that did not train on $i$; the baseline $\log_2\hat{p}(y_i)$ is the empirical label prior computed on that fold’s training split. Example comments with their PVI values by stratum are provided in Appendix Table \ref{tab:pvi_examples_gray} and Table S1 in the supporting information (SI).

\subsection{LLM Inference}
We run LLMs to predict \emph{approve}/\emph{remove} from a subreddit’s name, description, rules, and the comment under review. Open-source models are executed on the Aktus AI NVIDIA GPU clusters; proprietary models are queried via vendor APIs. We evaluate four open-source models (\emph{Llama 8B}, \emph{Qwen 7B}, \emph{Llama 70B}, \emph{Qwen 32B}) and two proprietary models (\emph{Gemini Flash} and \emph{GPT-5 Mini}). We use the checkpoints listed in Table~\ref{tab:model-variant}. We design eleven prompt templates (see Tables S2 and S3 in the SI) spanning: binary-answer formats, open-ended justifications, structured JSON outputs, gently action-biased variants, and role-specific instructions (e.g., ``experienced moderator’’ vs.\ ``generic annotator’’). We apply a lightweight post-processor to recover a binary decision, and when \emph{remove} is chosen, the violated rule index. Outputs that simultaneously indicate both actions (approve, remove) are marked \emph{conflicting}; responses from which no decision can be reliably extracted are \emph{unknown}. We use \emph{coverage} to denote the fraction of responses that yield a valid unambiguous action label.  

\begin{figure*}
  \centering
  \includegraphics[width=.7\linewidth]{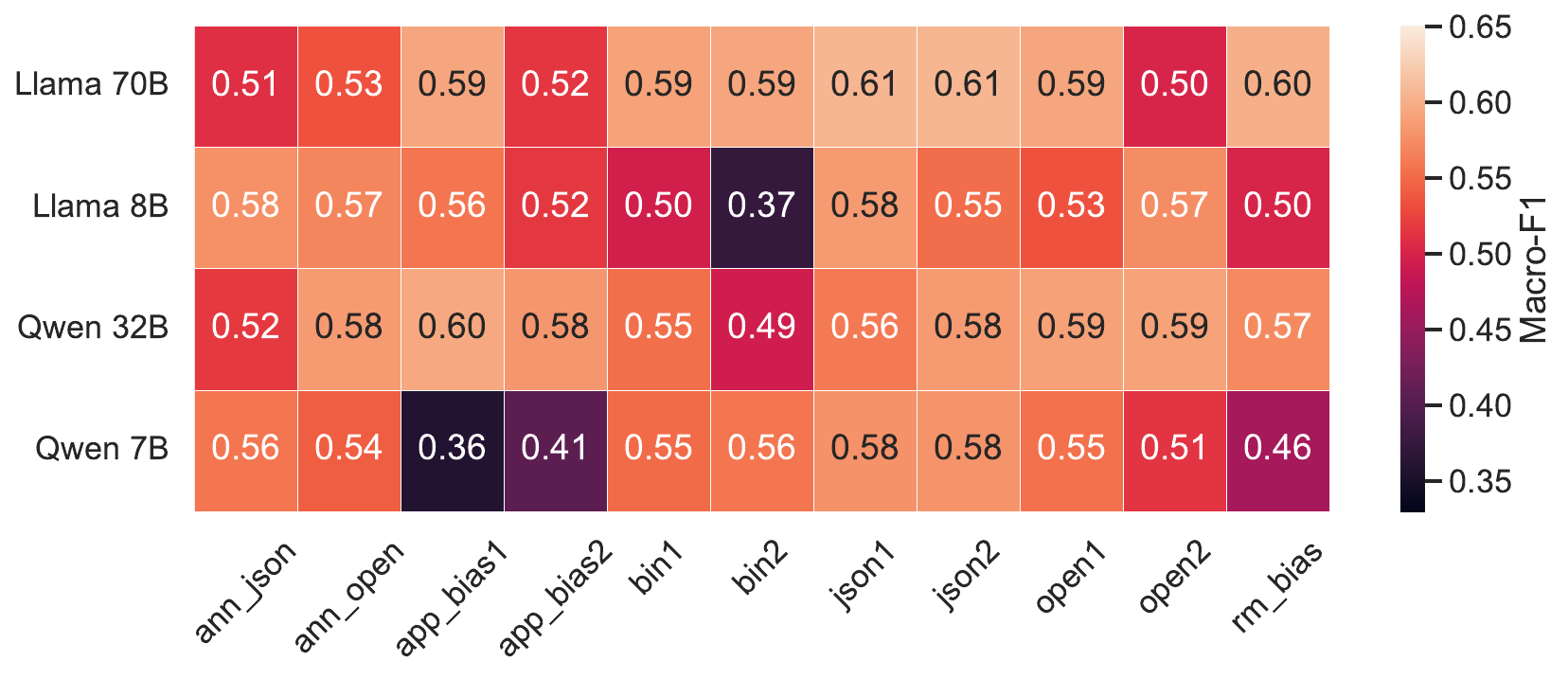}
  \caption{Heatmap of macro F1 scores comparing model responses to final moderation labels across prompt variations. JSON template prompts achieve the highest performance in three out of four language models and show the best performance consistency across models}
  \label{fig:prompt_pooled_f1}
\end{figure*}

\paragraph{Prompt variants.}
Across 11 prompt variants tested on open-source models, JSON-output templates achieve the highest coverage (SI, Figure S1) and in 3 out 4 models, the highest macro-F1 across models (Appendix Figure~\ref{fig:prompt_pooled_f1}). We also noticed that the prompt sensitivity decreases with models scale. For example, the standard deviation of macro-F1 across prompts drops from $\approx 0.073$ in \emph{Qwen 7B} to $\approx 0.032$ in \emph{Qwen 32B}. Based on these results and to contain API costs, we use the \texttt{json2} template when querying the proprietary model APIs.

\begin{table}[!htbp]
\setlength{\tabcolsep}{0.75mm}
\small
\sffamily
\centering
\begin{tabular}{l|l}
\toprule
\textbf{Short name} & \textbf{Checkpoint / Version} \\
\midrule
Llama 8B        & \texttt{meta-llama/Llama-3.1-8B-Instruct} \\
Qwen 7B         & \texttt{Qwen/Qwen2.5-7B-Instruct} \\
Llama 70B       & \texttt{meta-llama/Llama-3.3-70B-Instruct} \\
Qwen 32B        & \texttt{Qwen/Qwen3-32B} \\
Gemini Flash    & \texttt{gemini-1.5-flash-002} \\
GPT5 Mini       & \texttt{gpt-5-mini-2025-08-07} \\
\bottomrule
\end{tabular}
\caption{Model checkpoints and their short names used throughout the paper.}
\label{tab:model-variant}
\end{table}

\begin{table*}[]
\centering\sffamily\small
\resizebox{\textwidth}{!}{%
\begin{tabular}{@{}p{2cm}p{14cm}p{1.5cm}p{2cm}@{}}
\toprule
\textbf{Stratum} & \textbf{Content} & \textbf{PVI} & \textbf{Final Action} \\
\midrule

\multirow[t]{6}{2.5cm}{\textit{Gray Human}}
  & Fuck your pedo sympathizing bullshit. Now give me my ban so i never have to look at this faggotry every again & 0.906 & \texttt{remove} \\
  \cmidrule(lr){2-4}
  & Why did the 2003 lawsuit transcripts cost over \$4,000? A little while ago, I tried to locate the depositions stemming from Avery's 1985 wrongful conviction and subsequent lawsuit. I was told, despite what had been advertised, that the depositions were only available to those who donated and that 100\% of those who donated were perfectly happy with how things were run, thank you very much. I was further instructed to... & 0.856 & \texttt{approve} \\
  \cmidrule(lr){2-4}
  & I had my doubts, but after the smoke cleared, it was a good experiment & -2.887 & \texttt{remove} \\
  \cmidrule(lr){2-4}
  & Glad to see we're on our way back up! Currently buying in my downtime at work & -2.412 & \texttt{remove}  \\
\cmidrule(lr){2-4}

\multirow[t]{6}{2.5cm}{\textit{Gray Bot}}
  & Yes, play it. I think it's great, but I think with any work it's best to check it out for yourself and form your own opinion on it. This particular line is from a ""backstage"" feature, which is more lighthearted and the characters speak freely about how they feel.  It's not really meant to be keeping in tone with the rest of the story, which is dark and bleak, and the script reflects that. If you have it on the PS4 or Vita, you won't even get this feature as it's not in that version.  I do think the backstage features in the first game and its prequels are great in their own right as they dive deeper into the characters and themes and the intent behind the story. & 0.945 & \texttt{approve} \\
  \cmidrule(lr){2-4}
  & It's not too late, *and* I'd recommend it only if you have chess-playing friends IRL. If there's a brick-and-mortar chess club or Meetup near you, drop by and see how you like it. Learning chess as a purely online endeavor will be difficult, and more importantly, not fun. In the US, chess is primarily played by male children and twenty-somethings, and then by retirees. The ""middle-aged chess player"" demographic - I am one - is unfortunately a bit sparse. & 0.938 & \texttt{approve} \\
  \cmidrule(lr){2-4}
  & THE GHOST OF A PATRIOT HAUNTS OUR ENEMIES: His name was Seth Rich, and he was murdered by a bitch, a whore and a witch, she will be tried and hanged and left in a ditch. \#jesuslovesyou & -2.744 & \texttt{approve} \\
    \cmidrule(lr){2-4}
  & God, I LOVE lichess and chess.com almost as much as chess24 and \textless insert competitor\textgreater. Fuck the mods. & -2.333 & \texttt{remove} \\
\\ \bottomrule
\end{tabular}%
}
\caption{Sample of cases categorized as Gray Human or Gray Bot, with corresponding PVI scores and final actions.}
\label{tab:pvi_examples_gray}
\end{table*}

\begin{table*}[!htbp]
\setlength{\tabcolsep}{4pt}
\footnotesize
\centering
\sffamily
\begin{tabular}{c|cc|cc|cc|cc}
\toprule
\multirow{2}{*}{\textbf{Model}} & 
\multicolumn{2}{c|}{\textbf{Undisputed Human}} & 
\multicolumn{2}{c|}{\textbf{Undisputed Bot}} & 
\multicolumn{2}{c|}{\textbf{Gray Human}} & 
\multicolumn{2}{c}{\textbf{Gray Bot}} \\
\cmidrule(lr){2-9}
& F1 & 95\% CI & F1 & 95\% CI & F1 & 95\% CI & F1 & 95\% CI \\
\midrule
GPT5 Mini     & 0.59 & [0.59, 0.60] & 0.39 & [0.37, 0.41] & 0.42 & [0.40, 0.43] & 0.56 & [0.55, 0.57] \\
Gemini Flash  & 0.60 & [0.59, 0.61] & 0.47 & [0.45, 0.49] & 0.46 & [0.44, 0.47] & 0.60 & [0.59, 0.61] \\
Llama 70B     & 0.61 & [0.60, 0.62] & 0.51 & [0.49, 0.53] & 0.51 & [0.49, 0.52] & 0.59 & [0.58, 0.60] \\
Llama 8B      & 0.60 & [0.59, 0.61] & 0.38 & [0.36, 0.40] & 0.39 & [0.38, 0.41] & 0.56 & [0.55, 0.58] \\
Qwen 32B      & 0.58 & [0.56, 0.59] & 0.51 & [0.49, 0.53] & 0.52 & [0.50, 0.53] & 0.56 & [0.55, 0.57] \\
Qwen 7B       & 0.60 & [0.59, 0.61] & 0.47 & [0.45, 0.49] & 0.46 & [0.45, 0.47] & 0.57 & [0.56, 0.58] \\
\bottomrule
\end{tabular}
\caption{Macro-F1 and 95\% confidence intervals for different models across case strata.}
\label{tab:f1-buckets}
\end{table*}

\begin{figure}
  \centering
  \includegraphics[width=.9\linewidth]{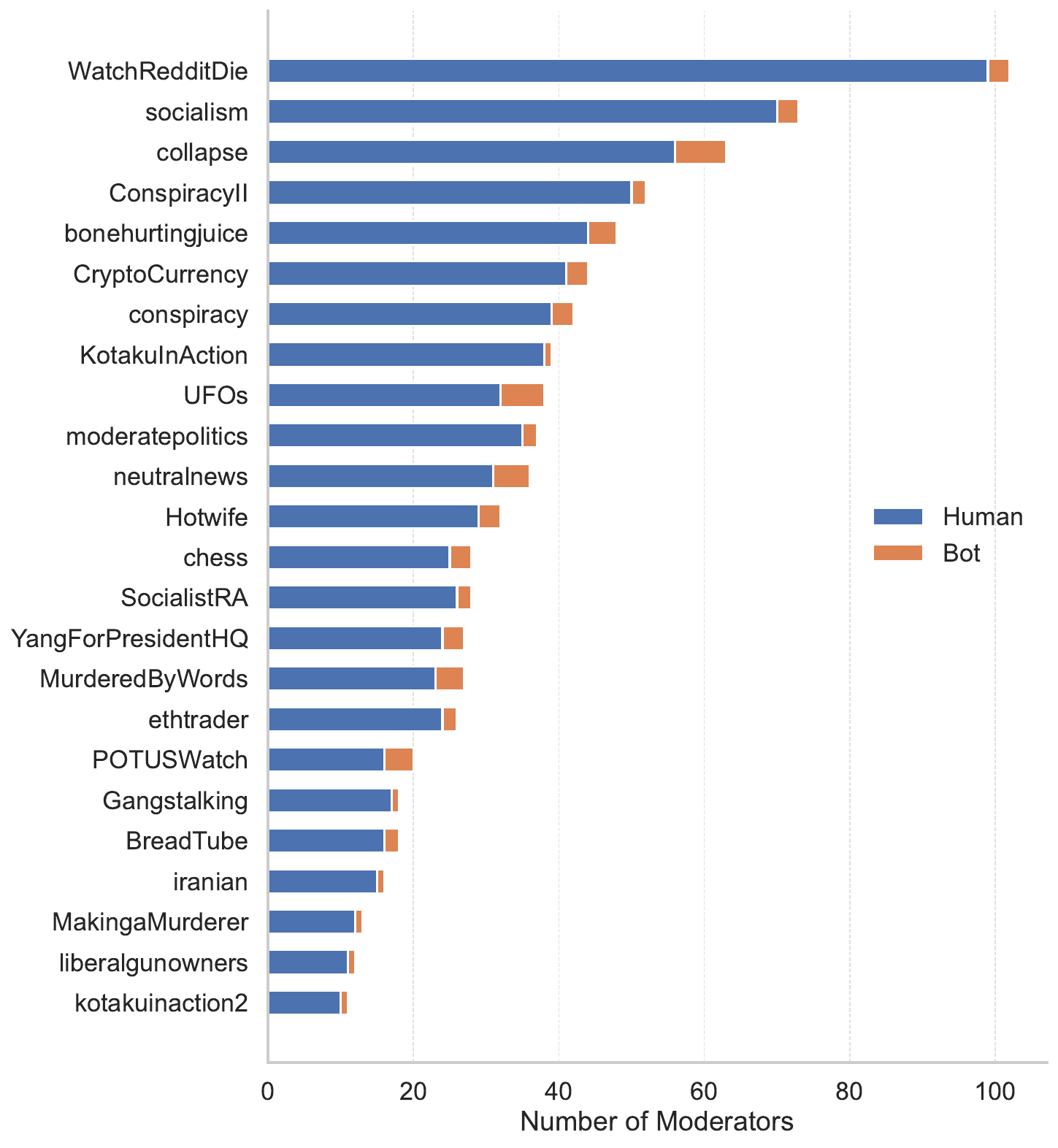}
  \caption{Number of unique human and bot moderators per subreddit in ModLog.}
  \label{fig:human_bot_mod_per_sub}
\end{figure}

\begin{figure}
  \centering
  \includegraphics[width=.9\linewidth]{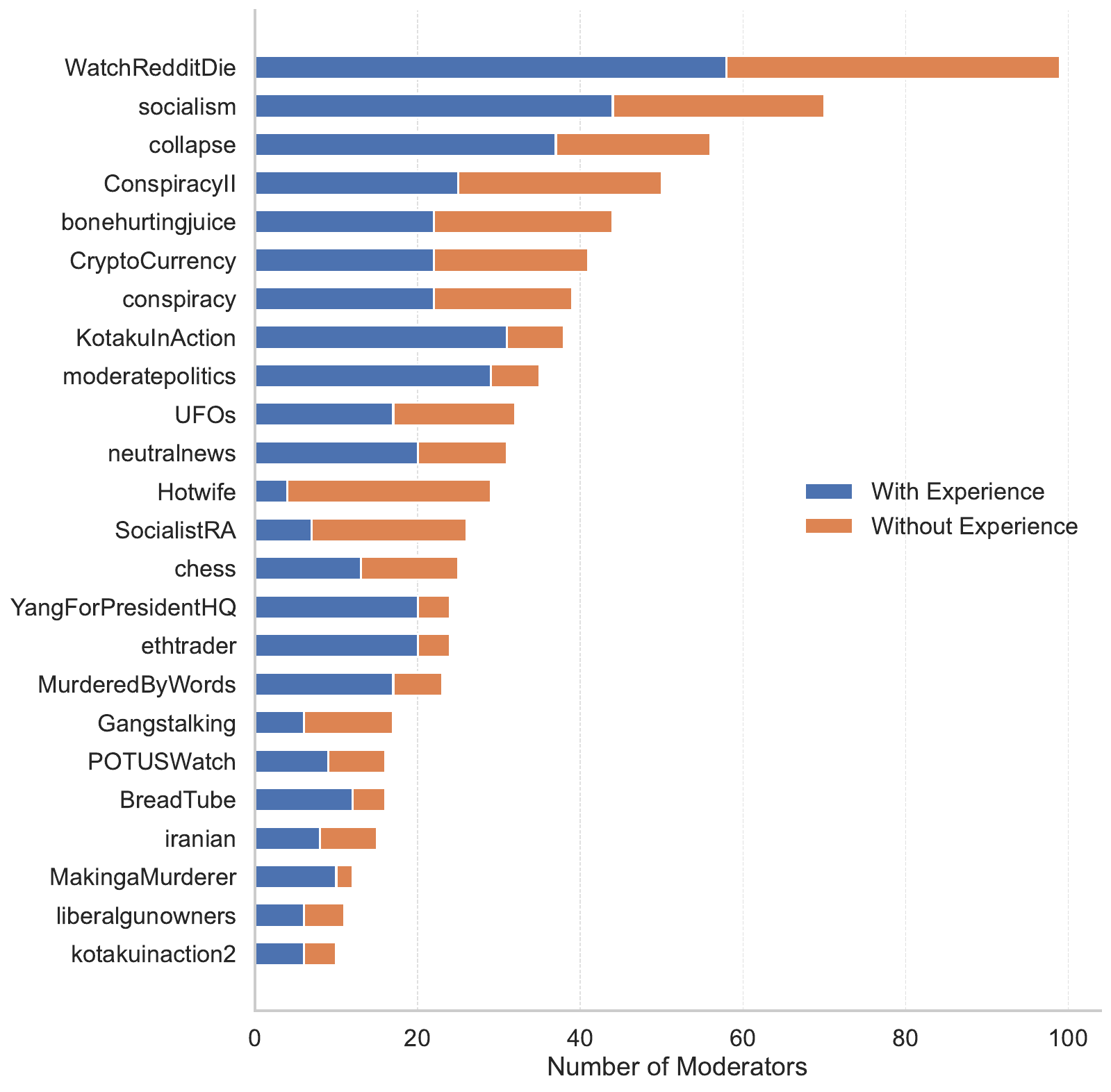}
  \caption{Number of human moderators with a valid joined date as moderators per subreddit compared to those without a valid joined date in ModLog.}
  \label{fig:mods_in_meta_per_sub}
\end{figure}

\end{document}